\documentclass[aps,prd,twocolumn,floatfix,superscriptaddress]{revtex4-2}
\usepackage{tabularx}
\usepackage{graphicx}
\usepackage{amsmath}
\usepackage{amssymb}
\usepackage{comment}
\usepackage{xcolor}
\usepackage{xspace}
\usepackage{multirow}
\usepackage{natbib}
\usepackage{hyperref}
\usepackage{subcaption}

\usepackage{subcaption}
\usepackage{cleveref}

\begin{document}
\title{A forest of gravitational waves in our Galactic Centre}

\author{Pau Amaro Seoane}
\affiliation{Universitat Politècnica de València, València, Spain}

\author{Shao-Dong Zhao}
\affiliation{Institute of Theoretical Physics \& Research Center of Gravitation,\\ Lanzhou University, Lanzhou 730000, China}
\affiliation{Lanzhou Center for Theoretical Physics, \\Key Laboratory of Theoretical Physics of Gansu Province, \\Key Laboratory of Quantum Theory and Applications of MoE, \\Gansu Provincial Research Center for Basic Disciplines of Quantum Physics, \\Lanzhou University, Lanzhou 730000, China}
\affiliation{Universitat Politècnica de València, València, Spain}

\begin{abstract}
At the Galactic Centre, we can expect a population of a few tens of early extreme-mass ratio inspirals (E-EMRIs) and extremely large mass ratio inspirals (XMRIs).
Depending on their evolutionary stage, they can be highly eccentric, with moderate signal-to-noise ratios (SNRs) of tens or hundreds, or nearly circular, with SNRs as large as a few thousand. Their individual signals combine into a common signal, which can complicate the resolution of other types of sources.
We have calculated the foreground signal of continuous E-EMRIs and XMRIs using a catalog based on the expected number of sources and a realistic phase-space distribution.  
The forest of E-EMRIs will cover a large portion of the LISA sensitivity curve, obscuring the signals of some massive black hole binaries, verification binaries, and harmonics of EMRIs in their polychromatic phase. The combined signal from XMRIs will be much weaker but still affect intermediate-mass black hole binaries. Due to the large SNR, this forest can be also found in other galactic nuclei, such as that of the Andromeda
galaxy.
Even under conservative assumptions, the forest created by E-EMRIs and XMRIs in our Galactic Centre will likely pose a challenge for resolving other types of sources, as their contribution is non-coherent and exhibits large SNRs.
\end{abstract}
\maketitle

\section{Introduction}

The gravitational capture of a compact object by a supermassive black hole (SMBH) represents one of the most interesting sources for a space-borne observatory like LISA \citep{LISA2017,AmaroLRR,Amaro2007}. Although the event rate of an extreme-mass ratio inspiral (EMRI) is very low for a single galaxy like the Milky Way (approximately $10^{-5}/\text{yr} - 10^{-6}/\text{yr}$), the cosmic event rate should yield a few detections per year, as we expect to observe them out to redshifts of $z\sim 1-4$ \citep{babak2020lisa,amaro2019extremely}.  

Recently, \cite{amaro2024mono} showed that we can expect a few tens of EMRIs in their early stages at the Galactic Centre. These early EMRIs (E-EMRIs) maintain a signal-to-noise ratio (SNR) of at least 10 up to half a million years before crossing the SMBH's event horizon. Given LISA's mission duration of a few years, E-EMRIs appear observationally as monochromatic sources, with their peak frequency fixed at a specific value—despite a cascade of harmonics spanning a wide frequency range. As they evolve in phase space, their peak frequency may shift toward higher values, transitioning them into an oligochromatic regime. Finally, polychromatic EMRIs are those weeks or months from merger, rapidly scanning a broad frequency range, as detailed in \cite{amaro2024mono}.  

On the other hand, substellar objects like brown dwarfs can approach the SMBH without undergoing strong tidal stresses, becoming powerful gravitational-wave sources in the extremely large mass ratio regime. In their late evolutionary stages, these sources achieve exceptionally high SNRs \citep{amaro2019extremely,gourgoulhon2019gravitational}.
Even in earlier stages, observing just one could enable precise parameter extraction for the Galactic Centre's SMBH, with minimal errors \citep{Veronica2023}.  

The solution derived in the works \cite{amaro2019extremely,amaro2024mono} is a steady-state one, meaning that as long as SgrA* fulfills the basic assumption of two-body relaxation and has a mass large enough so as to be considered to be at the centre of the stellar system (i.e. it does not wander off), which corresponds to the range of masses $10^{5}\,M_{\odot}-\text{a~few}\,10^{7}\,M_{\odot}$, there will permanently be a few tens of E-EMRIs and XMRIs lurking at the Galactic Centre.

In this work, we present the collective signal from E-EMRIs and XMRIs at the Galactic Centre, using a realistic source catalog and conservative assumptions. We show that the the cumulative forest of weaker sources can obscure other signals such as massive binaries or individual, polychromatic EMRIs, as well as verification binaries. Since this forest is a superposition of weak, independent sources with random phases, it is noise-like but non-Gaussian (unlike instrumental noise).  Unlike a stationary stochastic background, this forest may have time-dependent features, since some of the sources might be oligochromatic and and gets modulated by the antenna pattern, since the sources are concentrated at the Galactic Centre. Moreover, while the combined background is incoherent, some individual sources may have SNRs high enough to bias noise estimates if not modeled \citep{amaro2019extremely,gourgoulhon2019gravitational}. 

This is the reason why we refer to this collective signal as ``forest''. The whole contribution will be a combination of background and foreground sources, since frequency drift ($\Delta f \ll 1/T_{\text{obs}}$), and high eccentricity, predominantly contribute to the background noise. Their dense spectral overlap and lack of resolvable frequency evolution cause them to blend into a stochastic confusion foreground, analogous to the unresolved Galactic binary population in LISA. In contrast, oligochromatic asymmetric binaries, with moderate abundance, higher SNR, measurable frequency drift ($\Delta f \sim 1/T_{\text{obs}}$), and lower eccentricity, occupy the foreground or, maybe more accurately, the midground regime. Their slight frequency evolution allows partial resolvability over long observations, but their collective emission may still appear as structured excess power before individual extraction. The transition between these classes depends on the observation time $T_{\text{obs}}$. As $T_{\text{obs}}$ increases, some monochromatic sources may shift into the midground if their cumulative SNR rises, while oligochromatic sources gradually become fully resolvable foreground sources, polychromatic ones. The empirical criterion for resolvability requires both $\text{SNR} > \rho_{\text{thresh}}$ (e.g., $\rho_{\text{thresh}} \approx 10$) and $\Delta f \cdot T_{\text{obs}} > 1$, ensuring separation from the background. For data analysis, this implies that the monochromatic population is treated as an incoherent noise component, while the oligochromatic family demands targeted searches or hierarchical subtraction to mitigate their impact on background estimates.

\section{Catalog and strain}

\subsection{Forest source catalog}

An XMRI system consists of a brown dwarf orbiting a supermassive black hole (SMBH). Brown dwarfs have masses ranging from approximately 13 to 80 Jupiter masses ($0.013-0.0764\,M_{\odot}$). We therefore set this range to $0.01-0.08\, M_{\odot}$ to ensure complete coverage of the masses. In contrast, an E-EMRI system features a stellar-mass black hole orbiting an SMBH. Based on LIGO/Virgo observations \cite{fishbach2017ligo,fishbach2021ligo}, we adopt a mass range of $10-100 \,M_{\odot}$ for the stellar-mass black hole component in these systems. Following \cite{amaro2019extremely,amaro2024mono}, we look at the Galactic Centre (GC), so that we fix the distance to $8.3\,\text{kpc}$, we set the mass of the SMBH to $4\times 10^6 M_\odot$.

The critical radius in EMRIs/XMRIs represents the semimajor axis at which the binary's evolution undergoes a fundamental transition. Above this threshold, external perturbations from the surrounding stellar environment dominate orbital decay through processes, i.e. two-body relaxation \cite{AmaroLRR,Amaro2007}. Below it, gravitational-wave emission becomes the primary driver of the inspiral. This boundary marks where the compact object becomes irrevocably bound to the SMBH, entering a deterministic evolutionary path toward merger with predictable gravitational-wave signatures. By adopting $a_{\mathrm{crit}}$ as the initial condition for orbital evolution calculations, we isolate the relativistically significant phase of these systems, ensuring physical consistency and computational efficiency in modeling their gravitational-wave emission. This threshold distance has been analytically derived in \cite{amaro2019extremely} for XMRIs and in \cite{amaro2024mono} for E-EMRIs.

\begin{figure*}[hbt]
    \centering
    \includegraphics[width=1\linewidth]{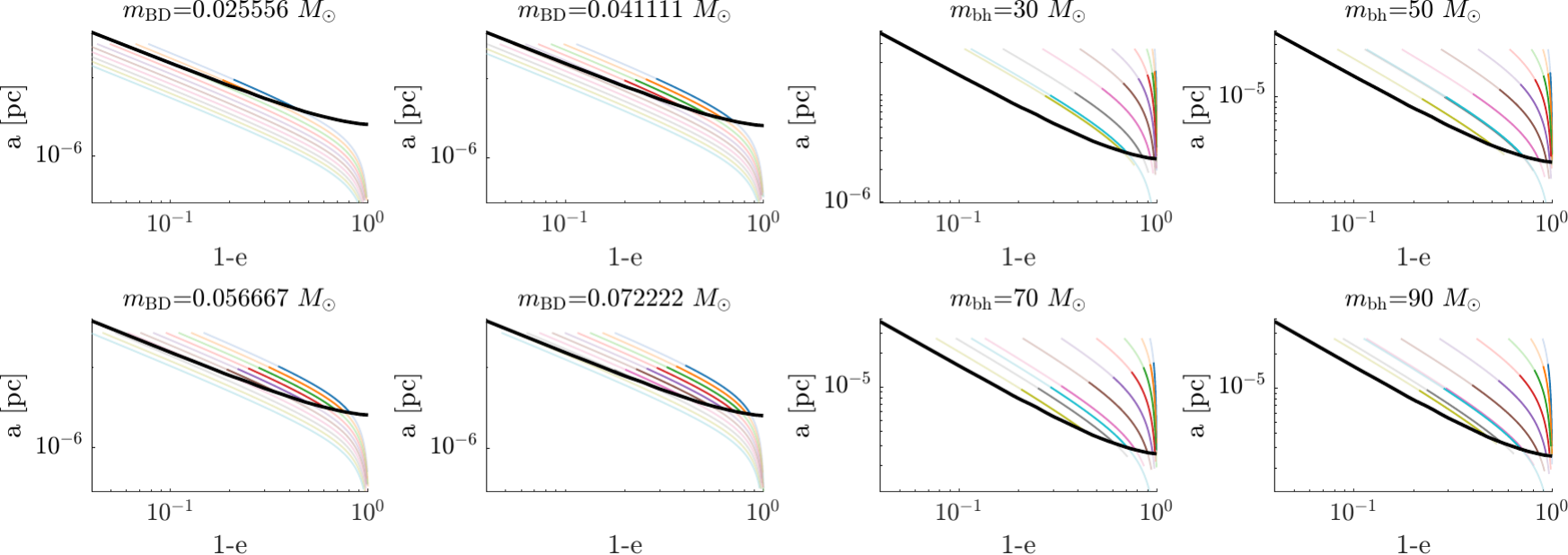}
    \caption{\textit{Left four panels:} Evolution in phase-space of a family of XMRIs with different initial parameters and 
    a given, fixed mass. The maximum value of the initial eccentricity corresponds to the top curve. We mark with a darker 
    colour the part of the evolution in which the source is entirely within the LISA frequency band. \textit{Right four panels:} 
    Same for a family of E-EMRIs. The thick solid line represents the approximate location of the LSO for the sources.}
    \label{fig:lso_orbit}
\end{figure*}

In Figs.~(\ref{fig:lso_orbit}), we show the orbit evolution track of XMRI/E-EMRI systems for different masses and different initial eccentricities when the orbits have a semi-major axis value of $a_{\mathrm{crit}}$, so that we will refer to the corresponding eccentricities as the critical eccentricity. The figures clearly show that for XMRI systems with low-mass brown dwarfs, only a small portion of the source's orbital evolution is fully in band.

We construct the source bank for XMRI/E-EMRI systems by independently sampling companion object masses within their respective ranges. Implementing a power-law mass distribution prioritizes higher-mass secondaries in our sampling. Critical orbital eccentricities are sampled logarithmically following the expected distribution of the works \cite{AmaroLRR,amaro2019extremely,amaro2024mono,amaro2013role}; while they initially are very high, by the time they have reached the LISA band, well after they have decoupled from the dynamics-dominated regime, they have reached lower values. These sampled masses and eccentricities, combined with the critical radii form the initial conditions for our orbital evolution calculations.

We calculate evolutionary trajectories for each unique initial configuration, secondary mass and critical eccentricity, using the approximation of \cite{peters1964gravitational}, in a re-arranged fashion,

\begin{align}
    \frac{df_{\rm orb}}{dt} =& \frac{96}{10\pi} \left(\frac{\mathcal{M}\, R_{\odot}} {c} \right)^{5/3}(2\pi f_{\rm orb})^{11/3} \nonumber \\
    &\times (1-e^2)^{-7/2}\left( 1+\frac{73}{24}e^2+\frac{37}{96}e^4\right) \label{eqs:dfdt} \\
    \frac{de}{dt} =& - \frac{e}{15} \left(\frac{\mathcal{M}\, R_{\odot}}{c} \right)^{5/3}(2\pi f_{\rm orb})^{8/3} \nonumber \\
    &\times (1-e^2)^{-5/2}\left( 304 + 121e^2\right), \label{eqs:dedt}
\end{align}

\noindent
where $\mathcal{M}$ represents the chirp mass of the binary system, defined as $\mathcal{M} = (m_1 m_2)^{3/5}/(m_1 + m_2)^{1/5}$, where $m_1$ and $m_2$ are the component masses. These individual trajectories are then combined to create complete XMRI and E-EMRI source banks used for the forest noise estimation.  

\subsection{Distinguishing between mono- and oligochromatic sources}

In order to calculate the forest, we need to estimate the signal amplitude of XMRI/E-EMRI sources that are in the LISA band. 
The rms (root mean square) strain of the $n$-th harmonic of inspiralling gravitational wave system is given by\cite{finn2000gravitational}

\begin{equation}\label{eqs:rms_strain}
    h_{\mathrm{rms},n} \equiv \sqrt{\langle h_{n+}^2+h_{n\times}^2\rangle}=\frac{1}{\pi 
     d f_{n}}\sqrt{\frac{G\dot E_n}{c^3}}
\end{equation}

\noindent
where $f_n=nf_{\rm orb}$. The root-mean-square strain per frequency bin represents the effective gravitational wave amplitude distributed across each harmonic of the orbital frequency. For a given harmonic $n$, the value $h_{\text{rms},n}$ quantifies how much the system's orbital energy loss $\dot{E}_n$ is converted into observable gravitational radiation at frequency $f_n = n f_{\text{orb}}$. This is analogous to measuring the brightness of light in a single pixel of a telescope image. Just as a pixel's intensity represents the combined light from all stars and diffuse sources within its angular resolution, $h_{\text{rms},n}$ captures the net gravitational wave amplitude from all orbital energy dissipated at a specific frequency bin. For background signals—like the unresolved glow of many faint stars contributing to a pixel's value—the rms strain statistically aggregates the incoherent superposition of weak gravitational wave sources within each frequency bin. In an unresolved gravitational wave foreground the phase information is lost but the collective power remains detectable. 

In order to separate the sources as monochromatic or oligochromatic (or polychromatic, but we are not considering these in the present work), we need then an argument based on the source's phase. The phase of the gravitational wave signal can be expressed using a Taylor expansion. Generally, we expand to the second order in $\varphi$, incorporating the first-order time derivative of frequency as:

\begin{equation}
    \begin{aligned}
        \varphi(t) &= \varphi_0+2\pi ft+\pi \dot{f}t^2\\
        &=\varphi_0+2\pi \left( f + \frac{1}{2} \dot{f}t\right) t
    \end{aligned}
\end{equation}

From the second line, we can define an effective frequency $f_{\mathrm{eff}}$ as:

\begin{equation}
    f_{\mathrm{eff}} \equiv f+\frac{1}{2}\dot{f}t
\end{equation}

This effective frequency $f_{\mathrm{eff}}$ differs from the instantaneous frequency $f$ by the product of the frequency derivative $\dot{f}$ and time $t$. The resulting phase difference between the instantaneous frequency and effective frequency is $\Delta \varphi=\pi\dot{f}t^2$. This phase difference increases quadratically with observation time, providing a distinctive signature that significantly enhances the detectability of gravitational wave signals against background noise. The quadratic phase evolution creates a unique spectral pattern that random noise processes cannot mimic consistently over extended observation periods, thereby improving our ability to extract individual signals using common data analysis techniques. As the compact object inspirals toward the central mass, its $n$th harmonic spends approximately $\sim f_n^2 / \dot{f}_n = \mathrm{d} \varphi_n / (2\pi\, \mathrm{d} \ln f_n)$ cycles near the frequency $f_n$. When a gravitational wave detector monitors the signal throughout the entire inspiral phase with an observational duration $T_{\mathrm{obs}}$ exceeding $f_n/\dot{f}_n$, the strain is enhanced by approximately the square root of this cycle count. This increase arises precisely because the phase coherence of the gravitational wave signal is maintained across the entire observation, while noise contributions tend to average out. Consequently, the resultant signal strength becomes approximately equivalent to that generated by a broad-band burst with the following amplitude

\begin{equation}\label{eqs:charateristic_strain}
     h_{c,n} \equiv h_{\mathrm{rms},n}\sqrt{\frac{2f_n^2}{\dot f_n}}=\frac{1}{\pi d}\sqrt{\frac{G\dot E_n}{c^3\dot f_n}},
\end{equation}

\noindent
which one refers to as the characteristic strain. This strain captures the total energy deposited across the frequency spectrum, analogous to how a short-duration burst of light spreads energy over a wide range of wavelengths. While the signal is narrow-band at any instant (monochromatic or oligochromatic), its frequency evolution over time causes it to sweep through a broad frequency range, resembling a burst when viewed over the entire observation. This distinguishes it from stationary sources (monochromatic signals) and highlights its detectability advantage; the phase coherence over many cycles boosts the SNR.

For a binary system in orbital motion that loses energy exclusively through gravitational wave emission, the rate of energy loss and the time derivative of the orbital frequency are given by the following expressions:

\begin{align}
    \dot E_n & = \frac{32 G^{{7}/{3}}}{5 c^5} (2\pi\mathcal{M}f_{\rm orb})^{{10}/{3}} g(n,e) \label{eqs:dEdt}, \\
    \dot f_{\rm orb} &= \frac{96 G^{{5}/{3}}}{5c^5} (2\pi)^{{8}/{3}} \mathcal{M}^{{5}/{3}}f_{\rm orb}^{{11}/{3}}\mathcal{F}(e) \label{eqs:dfdt}
\end{align}

In these equations, the functions $g(n,e)$ and $\mathcal{F}(e)$ contain the dependence on the harmonic number $n$ and orbital eccentricity $e$, accounting for the spectral structure.

The root-mean-square strain amplitude $h_{\mathrm{rms},n}$ can be derived by substituting the energy derivative from Eq.~\ref{eqs:dEdt} into Eq.~\ref{eqs:rms_strain}, 

    \begin{align}
    h_{\mathrm{rms}}^{2} & = \frac{1}{\pi^2 
     d^2 f_{n}^2}\frac{G\dot E_n}{c^3}=\frac{G \dot E_{n}}{c^{3}\pi^{2}d^{2}f_{n}^{2}}\\
     & =\frac{2^{10/3}32(G\mathcal{M})^{\frac{10}{3}}(\pi f_{\rm orb})^{\frac{4}{3}}g(n,e)}{5n^{2}d^{2}c^8}
    \end{align}

For the characteristic strain $h_{c,n}$, we need to include the frequency derivative from Eq.~\ref{eqs:dfdt},

    \begin{align}
    h_{c,n}^2 & =h_{\mathrm{rms},n}^2\frac{2f_n^2}{\dot f_n} =  \frac{1}{\pi^2 d^2}\frac{2G\dot E_n}{c^{3}\dot f_n} \\
     & =\frac{2(2\pi)^{{2}/{3}}(G\mathcal{M})^{{5}/{3}} g(n,e)}
     {3c^{3} f_{n}^{{1}/{3}} \mathcal{F}(e) \pi^2 d^2}
    \end{align}

For gravitational wave sources undergoing significant orbital evolution, we must weight their contribution by the maximum number of cycles observable at a given frequency during the observation period. This weighting is expressed as

\begin{equation}
    h_{c,n}\frac{1}{\mathcal{N}_{\mathrm{cyc,\,max}}}=h_{c,n}\frac{1}{f_nT_{\rm obs}}
\end{equation}

\noindent
where $\mathcal{N}_{\mathrm{cyc,\,max}} = f_nT_{\rm obs}$ represents the maximum number of cycles that can be observed at frequency $f_n$ during an observation time $T_{\rm obs}$. This weighting accounts for the fact that rapidly evolving sources spend less time emitting at any specific frequency, thus contributing proportionally less to the forest at that frequency.
The characteristic strain quantifies the total SNR of an inspiraling source by integrating its coherent phase evolution over the entire inspiral, effectively capturing its burst-like behavior. However, when calculating the forest contribution in a specific frequency bin, we must account for the limited time each source spends emitting within that bin during the observation period $T_{\mathrm{obs}}$. This is achieved through the weighting factor, which normalizes the source's contribution based on how many cycles it completes in that bin. While $h_{c,n}$ represents the cumulative detectability of the full inspiral, the weighting reflects our measurement constraints - rapidly evolving sources sweep through frequency bins too quickly to dominate any single bin, whereas slowly evolving or monochromatic sources concentrate their power in fewer bins. This distinction ensures the forest accurately represents what is resolvable within our finite observational window and frequency resolution, separating the intrinsic source strength from our technical limitations in resolving transient signals.  

The total gravitational wave forest (GWF) must therefore be calculated differently depending on whether sources are non-evolving/slowly-evolving or significantly evolving within the observation time. The comprehensive expression for the gravitational wave forest power spectral density becomes:

    \begin{align}
    &h^2_{\mathrm{gwf}}(f) = \\
    &\begin{cases}
        \int \mathrm{d} \mathcal{M} \mathrm{d} e
        \left[ \sum_n \frac{h^2_{\mathrm{rms},n}(f) ~ \mathrm{d}^3N }{ \mathrm{d} \mathcal{M} \, \mathrm{d} e \, \mathrm{d} \ln{f_{\rm orb}}}  \right], \dot f_n \,T_{\rm obs} < f_n \nonumber \\
        \int \mathrm{d} \mathcal{M} \mathrm{d} e
        \left[ \sum_n \frac{ {h^2_{c,n}(f)}/{[2f\,T_{\rm obs}}] ~\mathrm{d}^3N }{ \mathrm{d} \mathcal{M} \, \mathrm{d} e \, \mathrm{d} \ln{f_{\rm orb}}} \right] , \dot f_n   \,T_{\rm obs} \geq f_n
    \end{cases}
    \end{align}
    
The upper case applies to sources that evolve slowly enough that their frequency change during the observation period is less than the frequency itself ($\dot f_n \,T_{\rm obs} < f_n$). For these sources, the root-mean-square strain $h^2_{\mathrm{rms},n}(f)$ is the appropriate measure of their contribution to the forest. The lower case applies to rapidly evolving sources where the frequency change during observation is comparable to or greater than the frequency itself ($\dot f_n \,T_{\rm obs} \geq f_n$). For these sources, as stressed, we use the characteristic strain scaled by the observation time factor ${h^2_{c,n}(f)}/{2f\,T_{\rm obs}}$.

In both cases, we integrate over the chirp mass $\mathcal{M}$ and eccentricity $e$ distributions, and sum over $n$ harmonics. The term ${\mathrm{d}^3N}/({ \mathrm{d} \mathcal{M} \, \mathrm{d} e \, \mathrm{d} \ln{f_{\rm orb}}})$ represents the number density of sources per unit chirp mass, eccentricity, and logarithmic frequency interval. This differential approach allows us to properly account for the full population of gravitational wave sources with their diverse physical parameters.

\section{Results}

\subsection{XMRI forest}

For the XMRI population analysis, we sample $N \approx 20$ inband sources from our previously constructed heterogeneous XMRI source catalog. This is a representative number of the sources found by \cite{amaro2019extremely}. An inband source is defined as satisfying two criteria: (1) the source maintains orbital parameters ensuring its gravitational wave emission remains within LISA's frequency sensitivity band ($0.01\,\mathrm{mHz}$ to $10\,\mathrm{mHz}$) throughout the entire observation period (i.e. it is fully contained in LISA), and (2) the semi-major axis $a$ is larger than the last stable orbit radius $R_{\mathrm{LSO}}$ of the supermassive black hole, satisfying $a > R_{\mathrm{LSO}}$. The first criterion guarantees continuous detectability by constraining both the apoapsis and periapsis of the orbital motion, while the second criterion prevents plunge events during the observation window. Also, because the tidal radius of the brown dwarf is very close to the LSO, or even within the event horizon of SgrA*, depending on its mass \citep{amaro2019extremely}, we do not need to worry about it since we are considering much larger distances. 

\begin{figure}[hbt]
    \centering
    \includegraphics[width=1\linewidth]{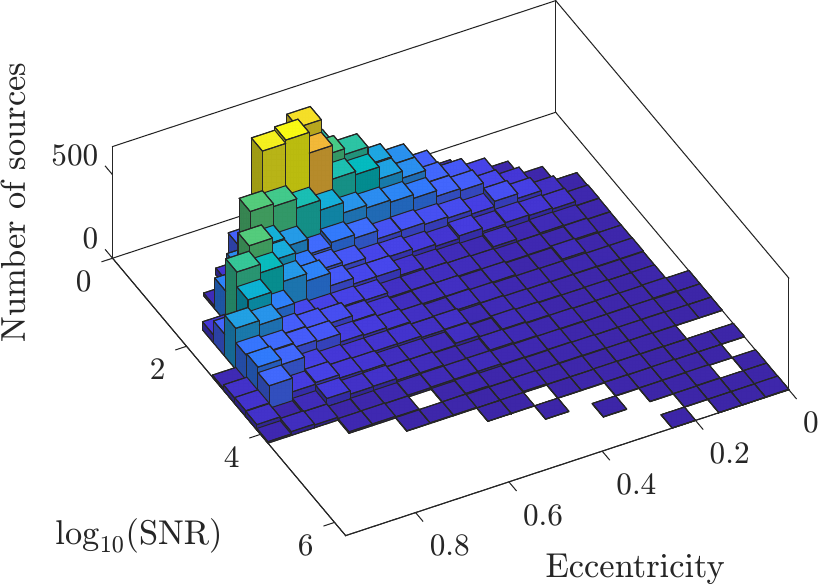}
    \caption{XMRI source bank in the eccentricity and SNR plane. We randomly extract the sources from this catalog. The distribution follows the results of the estimate of \cite{amaro2019extremely}. All sources are contained in the band of the detector.}
    \label{fig:2d_hist_XMRI}
\end{figure}

\begin{figure}
    \centering
    \includegraphics[width=1\linewidth]{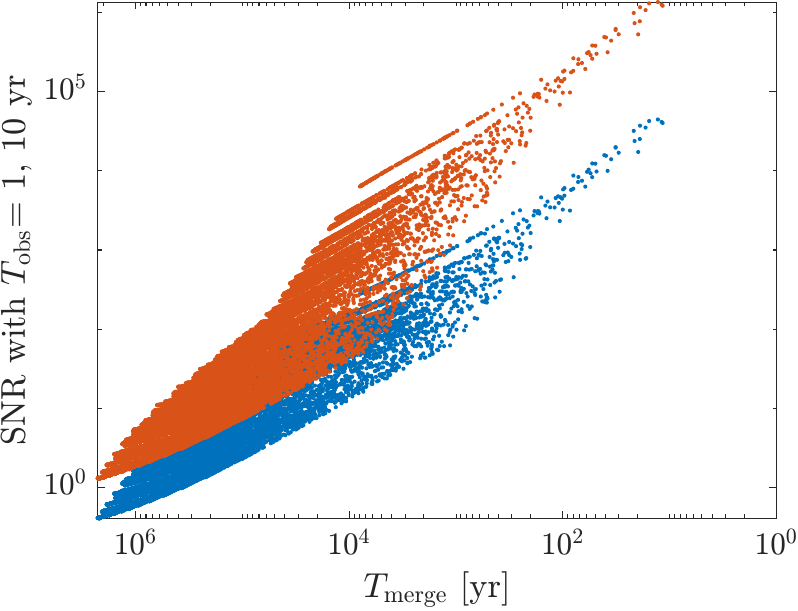}
    \caption{SNR as a function of $T_{\mathrm{merge}}$ for the complete XMRI catalog sources with different observation time, 1 year (blue dots) and 10 years (orange dots).} 
    \label{fig:SNR_Tmerge_XMRI} 
\end{figure}

Figure~\ref{fig:2d_hist_XMRI} shows the distribution of XMRI sources in the eccentricity–SNR plane, constructed from a randomly extracted sample consistent with the estimates of~\cite{amaro2019extremely}. Most detectable sources cluster at low SNRs $\mathrm{SNR} < 100$ and moderate to high eccentricities $e \sim 0.6-0.8$. 
Figure~\ref{fig:SNR_Tmerge_XMRI} further presents the relationship between SNR and the time to merger $T_{\mathrm{merge}}$ for the XMRI population, comparing 1-year (blue points) and 10-year (orange points) observation periods. As expected, longer observation times significantly enhance the accumulated SNR, especially for systems with shorter $T_{\mathrm{merge}}$.

We employ a stratified sampling approach to select sources across distinct SNR intervals to ensure balanced representation across the E-EMRI population. For a 1-year observation period, we draw $n_1 = 10$ sources from the low-SNR regime, defined as $\mathcal{S}_1 \in [10, 50]$, $n_2 = 8$ sources from the moderate-SNR range, $\mathcal{S}_2 \in [50, 500)$, $n_3 = 2$ sources from the intermediate-SNR range, $\mathcal{S}_2 \in [500, 1000)$ . This selection strategy enables us to capture both the weaker and moderately strong sources that contribute to the unresolved background. 

To quantify the maximum potential contribution of this population to the forest, we introduce an estimation procedure that calculates the cumulative normalized level across multiple realizations.

\begin{equation}
    V_{\mathrm{norm}} = \sum_i \sum_s f_i\, h_{\mathrm{bkg},\,s}(f_i),
\end{equation}

\noindent 
where $s$ corresponds to the individual forest sources, $i$ denotes the frequency bin index, $f_i$ represents the central frequency of bin $i$, and $h_{\mathrm{bkg},s}(f_i)$ is the strain amplitude of source $s$ at frequency $f_i$. We perform $N_{\mathrm{real}} = 10^3$ independent realizations to obtain a statistical representation of $V_{\mathrm{norm}}$.

\begin{figure}[hbt]
    \centering
    \begin{subfigure}{0.9\linewidth}
        \centering
        \includegraphics[width=\linewidth]{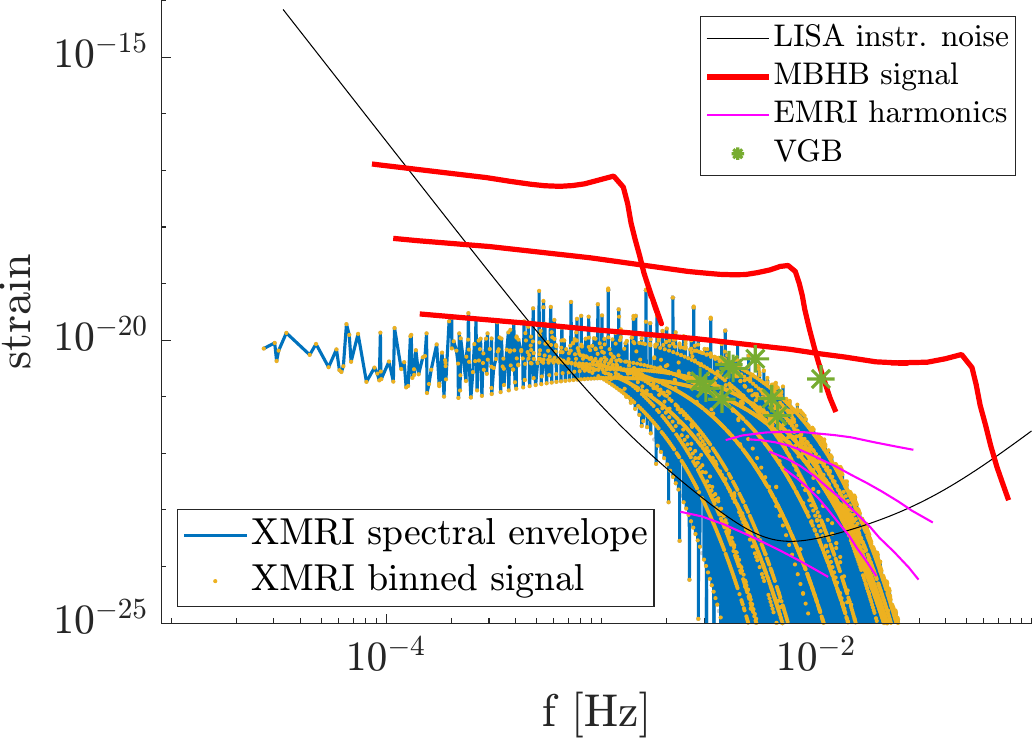}
        \caption{1-year XMRI forest configuration}
        \label{fig:XMRI_1yr_forest}
    \end{subfigure}
    
    
    \begin{subfigure}{0.9\linewidth}
        \centering
        \includegraphics[width=\linewidth]{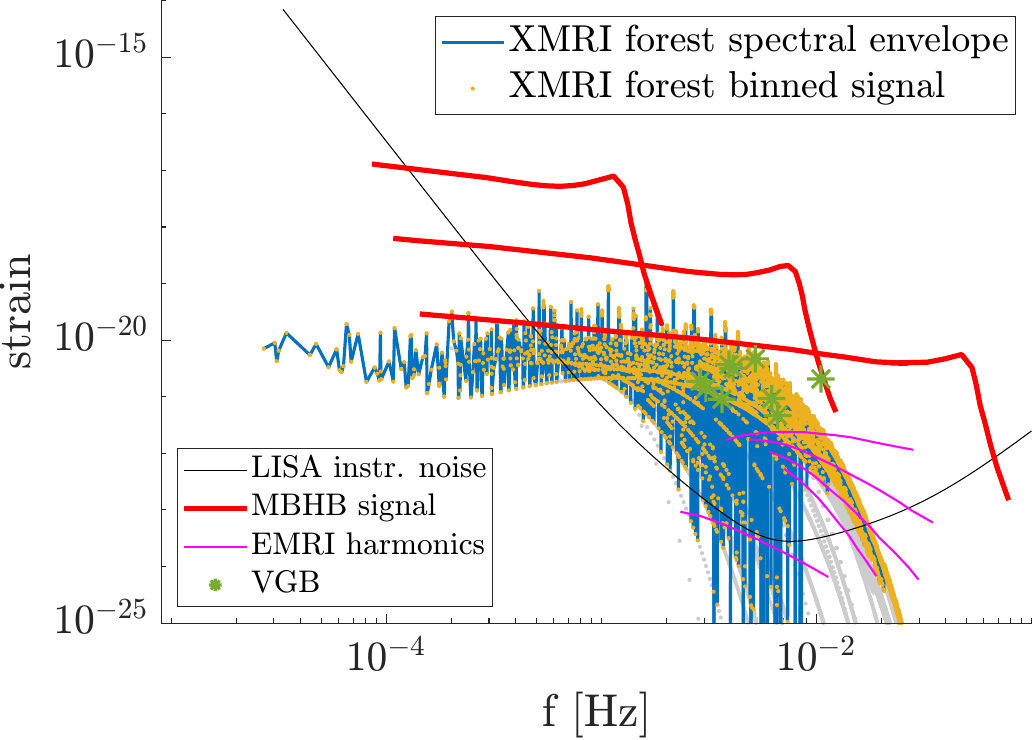}
        \caption{10-year XMRI forest configuration}
        \label{fig:XMRI_10yr_forest}
    \end{subfigure}
    
    \caption{Spectral characterization of the XMRI forest for observation periods of 1 year (top panel) and 10 years (bottom panel). Grey points indicate individual harmonic strains from distinct sources. Yellow dots represent the superposition of these individual strains within frequency bins, while the blue curve is obtained by connecting all non-zero strain values, illustrating the overall shape of the  spectrum. For comparison, representative signals from massive black hole binaries (red), Galactic verification binaries (green stars), EMRI harmonics (magenta),  LISA noise (black) are overlaid.}
    \label{fig:XMRI_forest}
\end{figure}

The results of our XMRI forest analysis are presented in Figure~\ref{fig:XMRI_forest}, along with several representative gravitational wave sources expected in the LISA band. Representative signals from massive black hole binaries , Galactic verification binaries, and characteristic EMRI harmonics are overlaid for comparison, together with the LISA instrumental sensitivity curve, consistent with those shown in Figure~(1) of \cite{LISA2017}. The yellow dots represent the superposition of individual XMRI strains binned in frequency, while the blue curve traces the non-zero strain values, illustrating the overall spectral shape of the forest. Notably, the XMRI background partially overlaps with the characteristic harmonics of polychromatic EMRI signals around frequencies of $\sim 1\,\mathrm{mHz}$, potentially introducing additional confusion noise in this region. Nevertheless, due to the low companion masses in XMRI systems, the background level remains relatively low and is not expected to substantially impede the detection of primary sources such as MBHBs or Galactic binaries.

It is important to note that the spectral forest is evaluated using different frequency resolutions corresponding to the two observation durations: 1 year and 10 years. Specifically, the frequency bin width is inversely proportional to the observation time, i.e., $\Delta f = 1/T_{\mathrm{obs}}$, resulting in finer resolution for the 10-year case. With high frequency resolution, individual source contributions become more distinguishable, appearing as sharp features or spikes in the spectrum. Despite this improvement in frequency resolution, the overall shape and amplitude of the forest remain nearly unchanged between the two durations. This consistency arises from the extremely slow dynamical evolution of XMRI systems, which leads to minimal frequency drift over the considered timescales. Nevertheless, in certain regimes—especially for the inspiral of low-mass MBHBs with total masses around or below $10^4\,M_{\odot}$—the XMRI forest could still introduce confusion noise or reduce signal clarity due to spectral overlap in narrow frequency bands.

\subsection{E-EMRI forest}

\begin{figure}[hbt]
    \centering
    \includegraphics[width=1\linewidth]{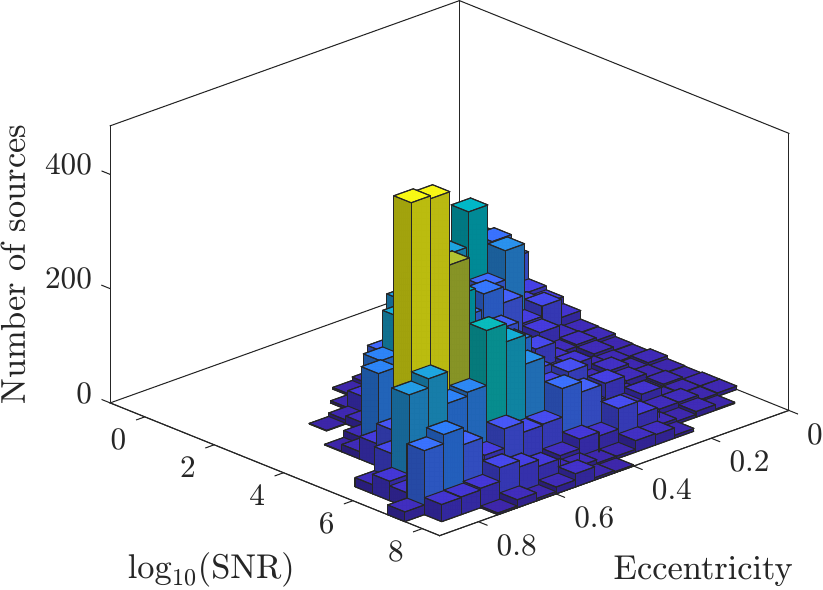}
    \caption{Same as Figure~(\ref{fig:2d_hist_XMRI}) but for E-EMRIs. The distribution follows the estimates of \cite{amaro2024mono}.}
    \label{fig:2d_hist_E-EMRI}
\end{figure}

\begin{figure}
    \centering
    \includegraphics[width=1\linewidth]{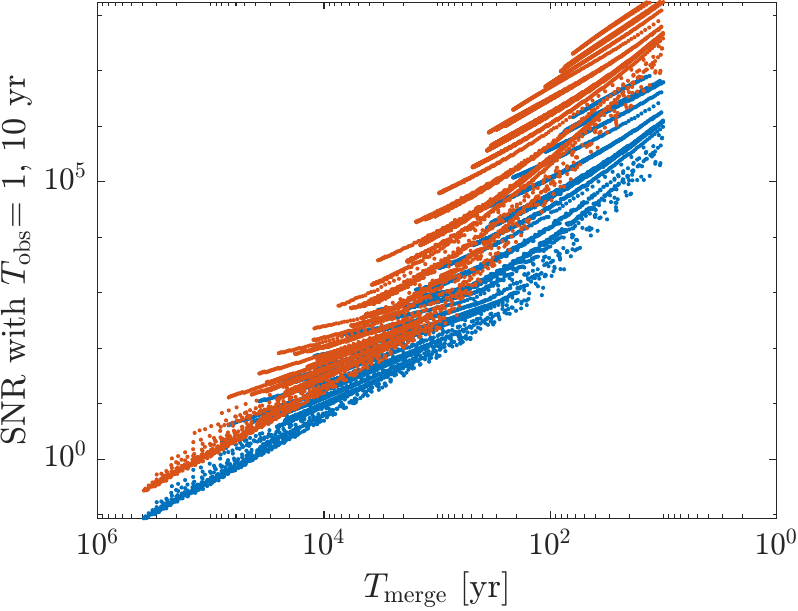}
    \caption{Same with Figure.~\ref{fig:SNR_Tmerge_XMRI} but with E-EMRI sources.}
    \label{fig:SNR_Tmerge_E-EMRI}
\end{figure}

For the E-EMRI population analysis, following the methodology outlined in \cite{amaro2024mono}, we sample approximately $N \approx 40$ inband sources from our E-EMRI source catalog. The SNR-eccentricity distribution of our catalog used in this sampling is illustrated in Figure~(\ref{fig:2d_hist_E-EMRI}). Similar to the XMRI analysis, we perform $10^3$ independent realizations. The sampling procedure for each realization utilizes a stratified approach across multiple SNR intervals to ensure representative coverage of the source population. For the 1-year observation period, we draw $n_1 = 25$ sources from the low SNR range $[10, 50)$, $n_2 = 10$ sources from the moderate SNR range $[50, 500)$, $n_3 = 4$ sources from the intermediate SNR range $[500, 1000)$, and $n_4 = 1$ sources from the high SNR regime ($\geq 1000$). In Figure.~\ref{fig:SNR_Tmerge_E-EMRI}, we show the SNR as a function of $T_{\mathrm{merge}}$ with different observation time as 1-year and 10-year.

\begin{figure}[h]
    \centering
    \begin{subfigure}[b]{0.9\linewidth}
        \centering
        \includegraphics[width=\linewidth]{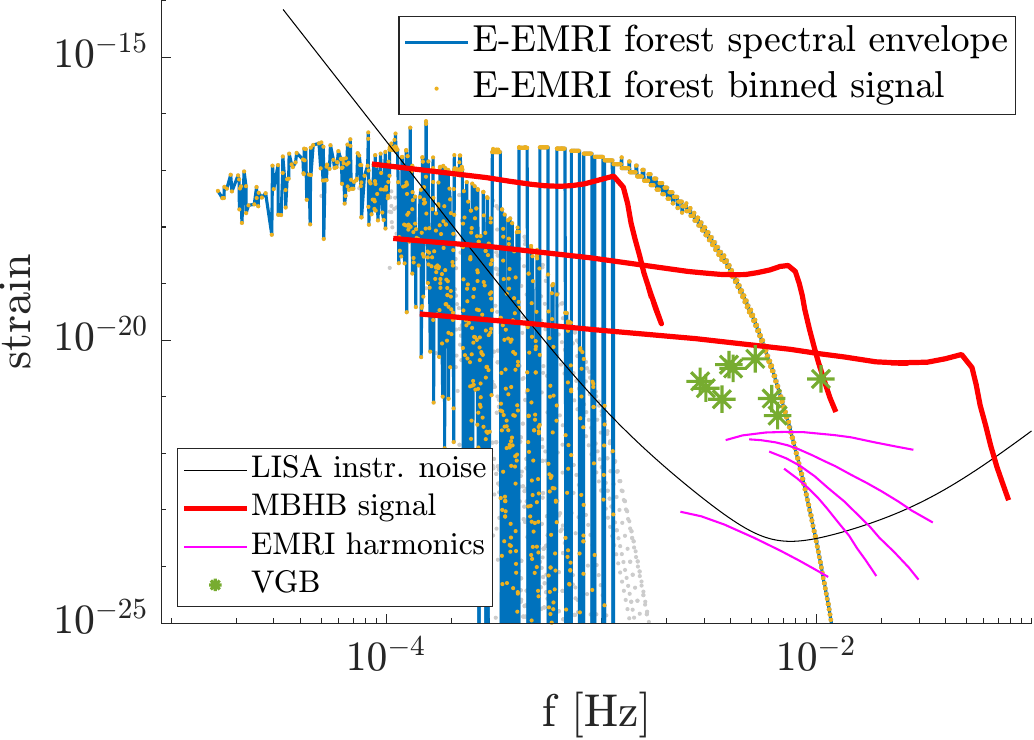}
        \caption{1-year E-EMRI forest configuration}
        \label{fig:E-EMRI_1yr_forest}
    \end{subfigure}
    
    
    \begin{subfigure}[b]{0.9\linewidth}
        \centering
        \includegraphics[width=\linewidth]{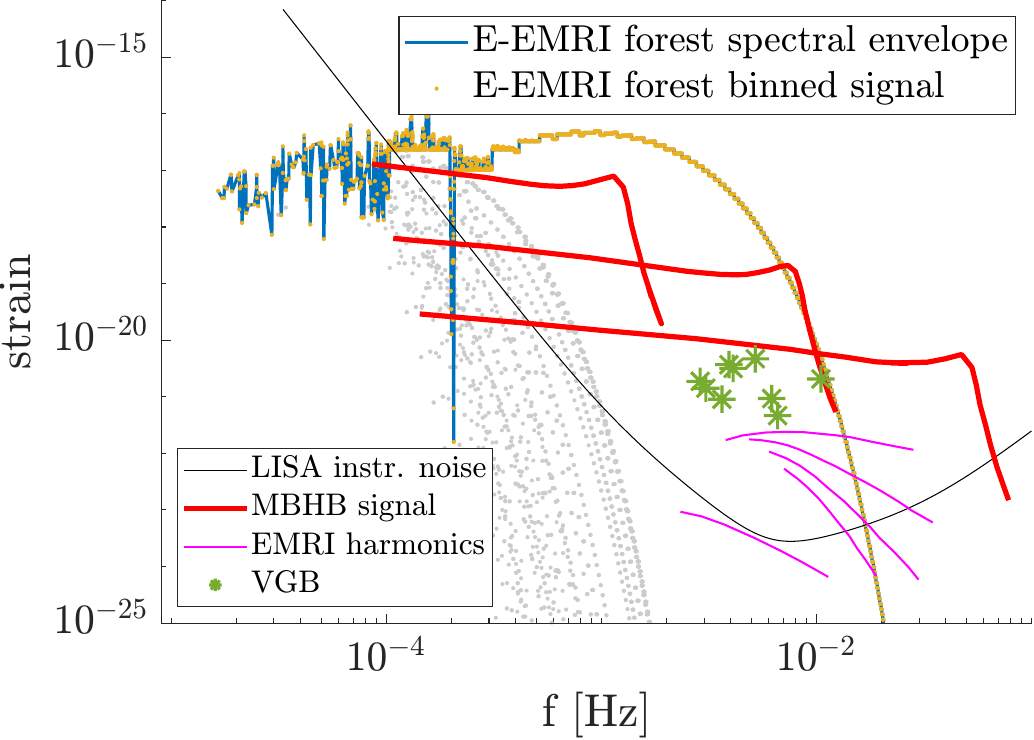}
        \caption{10-year E-EMRI forest configuration}
        \label{fig:E-EMRI_10yr_forest}
    \end{subfigure}
    
    \caption{Same as Figure~(\ref{fig:XMRI_forest}) but for the E-EMRI forest configurations over $T = 1$-year (upper panel) and $T = 10$-year (lower panel) observation periods. As in the other figure, we depict a few typical LISA sources.}
    \label{fig:E-EMRI_forest}
\end{figure}

Figure~\ref{fig:E-EMRI_forest} shows the forest contribution from E-EMRI sources in our mixed catalog. These sources exhibit high strain amplitudes, producing a pronounced forest in the LISA band. The visible harmonic structure arises from their eccentric orbits, which generate gravitational radiation bursts across multiple harmonics. For the 10-year observation (lower panel), the improved frequency resolution leads to a finer and smoother spectral structure, particularly at higher frequencies $f \gtrsim 1\,\mathrm{mHz}$. Compared to the 1-year case, the extended observation reduces confusion at low frequencies and sharpens the forest contribution at high frequencies, in particular in the region where we expect polychromatic EMRIs to populate phase-space.

The blue curve represents the overall shape of the forest spectrum, constructed by connecting all non-zero strain values. At lower frequencies, the blue curve becomes increasingly irregular, exhibiting pronounced spectral spikes. This behavior arises primarily from the discrete harmonic structure of the sources and the limited number of contributing systems in this frequency range, which leads to uneven spectral coverage. Additionally, the high frequency resolution used in the analysis means that some frequency bins receive contributions only from harmonic of low SNR sources, which typically have lower strain amplitudes. As a result, certain bins may exhibit a sudden drop in the forest level, contributing to the jagged appearance of the curve. At frequencies above 1 mHz, the curve closely follows the distribution of yellow dots, which represent non-zero data points of the forest strain. In this high-frequency regime, the spectrum is dominated by high-order harmonics of a few very high SNR sources, leading to a relatively smooth and elevated strain profile. These harmonics fall within LISA's most sensitive band and are well-resolved due to the long observation duration. 

The distribution of individual EMRI source strain amplitudes (grey dots) exhibits a distinct bifurcation, forming two visibly separated branches. This separation arises from the variation in source SNR. The lower branch is predominantly composed of sources with intermediate SNRs, whereas the upper branch originates from a small subset of loud sources. Owing to our background construction strategy—which maximizes spectral coverage by selecting realizations with the densest frequency occupation—the upper envelope is shaped by extreme cases, including sources reaching high SNRs, which contribute to the strain spectrum at high frequencies, leading to a visibly elevated branch relative to the bulk of the population.

\subsection{Extragalactic forests: The Andromeda galaxy}

Following our investigation of Galactic E-EMRI signals, we extend our analysis to neighbouring galaxies to explore the detectability of these forests. Among potential targets, Messier 32 (M32)—a compact dwarf elliptical satellite of Andromeda—presents an ideal candidate due to its central supermassive black hole with mass $1.5\text{--}5\times10^6\,M_\odot$, comparable to Sgr A*, and its relatively modest distance of $760\,\mathrm{kpc}$ \citep{M322004}.

\begin{figure}[h]
    \centering
    \begin{subfigure}[b]{0.9\linewidth}
        \centering
        \includegraphics[width=\linewidth]{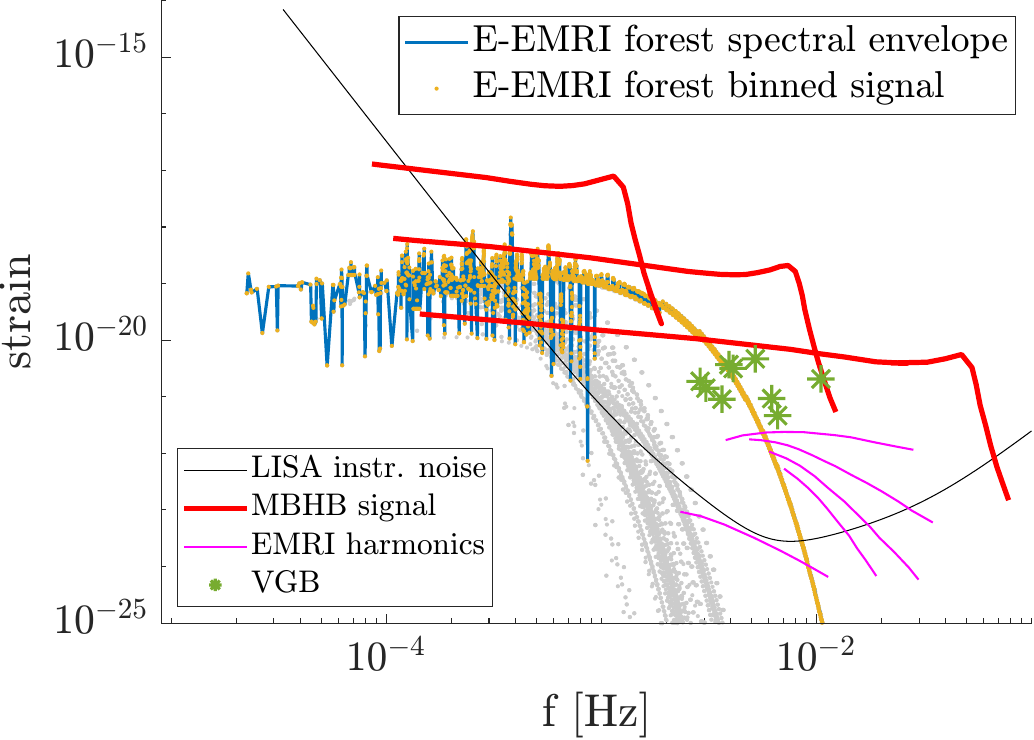}
        \caption{1-year E-EMRI forest configuration}
        \label{fig:E-EMRI_1yr_forest_agm}
    \end{subfigure}
    
    
    \begin{subfigure}[b]{0.9\linewidth}
        \centering
        \includegraphics[width=\linewidth]{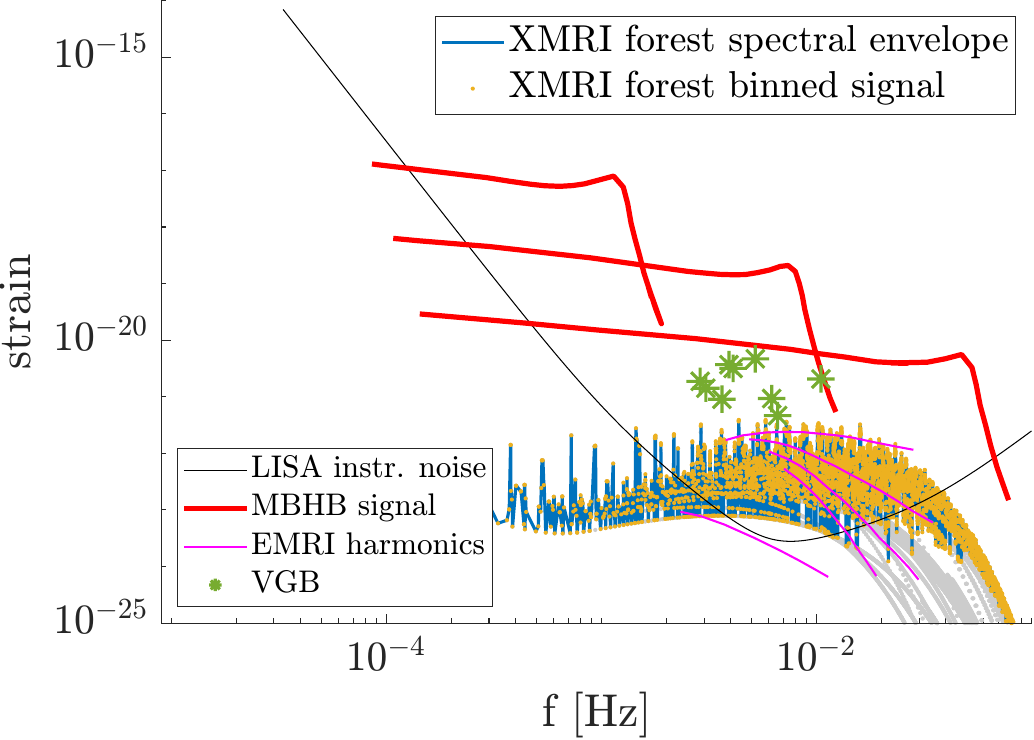}
        \caption{1-year XMRI forest configuration}
        \label{fig:XMRI_1yr_forest_agm}
    \end{subfigure}
    
    \caption{Characteristic strain versus frequency for potential EMRI sources from M32 (d $\approx$ 760 kpc). Upper panel shows E-EMRI signals, many remaining above LISA's sensitivity curve (red lines) in the $10^{-4}$--$10^{-3}$ Hz band. Lower panel shows XMRI signals with greater attenuation due to distance, with most falling below detection threshold. Colored curves represent various other LISA source types for comparison.}
    \label{fig:source_forest_agm}
\end{figure}

Figure~\ref{fig:source_forest_agm} shows characteristic strain spectra for both E-EMRI and XMRI populations from M32 over a 1-year observation period with a handful of sources, as an illustration. The upper panel demonstrates that despite the substantial distance, many E-EMRI signals remain above LISA's sensitivity curve in the $10^{-4}$ to $10^{-3}$ Hz band, particularly those involving higher-mass compact objects on highly eccentric orbits. In contrast, the lower panel shows that standard XMRI signals experience more severe attenuation, with most falling below the detection threshold across the frequency spectrum, although with a clear impact on polychromatic EMRIs. Obviously, understanding the shape of the forest when taking into account all close-by galaxies is important, and this will be addressed in a separate work.

\section{Conclusions}

 In this work, we characterize the gravitational wave forest signal originating from our Galactic Centre, generated by the populations of XMRIs and E-EMRIs. Building upon the steady-state solutions derived in \citep{amaro2019extremely,amaro2024mono}, the resulting GW forests presented here similarly represent steady-state configurations. The forest signals shown in Figures~(\ref{fig:XMRI_forest}) and (\ref{fig:E-EMRI_forest}) correspond to what LISA would observe both immediately upon launch and a thousand years into the future, reflecting the persistent nature of these sources.

 The combined signal from both families is termed a ``forest'' because it resembles a dense thicket of overlapping sources with varying resolvability. The monochromatic sources - numerous, faint, and static in frequency-form the background, similar to an undergrowth. The oligochromatic ones—fewer, brighter, and slowly drifting—act as the foreground, like distinguishable trees. The intermediate cases, where sources are partially separable but not yet fully resolved, constitute the midground, analogous to shrubs emerging from the underbrush. Together, these layers create a structured but entangled spectral landscape, justifying the metaphor of a gravitational wave ``forest''. The term captures the hierarchical density and observational challenge: just as a forest blends individual elements into a collective whole, the asymmetric binary populations merges into a composite signal requiring careful decomposition.

Our analysis reveals distinct impacts for the two populations: While the XMRI forest shows little influence on most typical LISA sources, it may affect binaries containing black holes with masses below $10^5\,M_{\odot}$. In contrast, the E-EMRI forest, owing to its more massive constituents, occupies a substantial portion of LISA's observational bandwidth. This forest obscures the inspiral, merger, and ringdown signals of very massive binaries ($\sim10^7\,M_{\odot}$) and significantly affects lower-mass binaries during their inspiral phase, with progressively stronger effects for more massive systems. Additionally, this forest obscures certain verification binaries and interferes with the detection of harmonics from polychromatic EMRIs.

When the frequency span of a given signal is smaller than the frequency resolution $f_{\text{res}} = 1/T_{\text{obs}}$, its presence remains detectable in the frequency domain, but its fine spectral features become unresolvable. In such cases, closely spaced frequency components within this narrow span merge into a single frequency bin due to the limited observation time, effectively blurring their distinct variations. This coarse frequency resolution prevents the discrimination of finer details, necessitating a longer $T_{\text{obs}}$ to achieve sufficient resolution. Even when the frequency shift is smaller than $f_{\text{res}}$, the signal remains observable provided its signal-to-noise ratio exceeds the detection threshold, though its spectral shape and other detailed characteristics cannot be resolved.

Removing the incoherent gravitational-wave background from sources like E-EMRIs and XMRIs in the Galactic Centre is a challenge due to its statistical nature and overlap with target signals. The difficulty arises from several factors: the background is a superposition of weak, independent sources with random phases, making it noise-like but non-Gaussian, unlike instrumental noise; unlike a stationary stochastic background, the forest we present in this work may exhibit time-dependent features since some of the individual sources are oligochromatic and hence there will be a frequency drift evolution at larger frequencies. Also, the Galactic gravitational wave foreground will exhibit strong non-stationarity mainly die spatial concentration towards the Galactic Centre, and modulation from LISA's antenna pattern evolution (see discussion of Galactic foregrounds in \cite{Digman2022}). While the full impact on parameter estimation requires detailed study through global fit analyses, current methodologies \cite{Littenberg2023,LittenbergEtAl2020} already demonstrate significant progress in foreground characterization. One should implement time-frequency domain modeling, where foregrounds are treated as non-stationary signals rather than quasi-stationary noise components. This approach naturally accommodates the expected spatial and temporal variability of both the Galactic background and resolvable EMRI populations. Additionally, the signals overlap in frequency with target sources, and while the combined forest is incoherent, some individual sources may have SNRs high enough to bias noise estimates if not modeled. Data analysis becomes complex because template-based subtraction fails for unresolved sources, and Bayesian hierarchical approaches are computationally expensive due to the high-dimensional parameter space (thousands of sources). 

Potential mitigation strategies include statistical subtraction by modeling the forest as a non-Poissonian noise process, which requires precise knowledge of the source population; multi-messenger constraints combining GW data with electromagnetic observations. Indeed, an XMRI is a brown dwarf revolving around SgrA* at extremely small peripasis distances, of a few Schwarzschild radii, and it is not hard to imagine that the flares that we observe at the Galactic Centre \citep{flares2003,von2025first,Ciurlo2025} have their origin here in particular because they are observed at a slightly shorter distance, of $2-4\,R_\text{Schw}$ \citep{genzel2024experimental}. The tidal stresses acting on to the brown dwarf are very likely the cause of such flares; this is an ongoing work which will be published elsewhere. These possibilities could allow us to break degeneracies in source counts. Also we could look into machine learning techniques to separate incoherent components from coherent signals; and global fit frameworks simultaneously fitting all resolvable sources and background \citep{LISAGlobal} (e.g., using LISA's global analysis pipelines). 
 
However, limitations remain, such as the computational cost of full marginalization over thousands of sources, and residual contamination from mis-modeled sources biasing parameter estimation for targets. While not impossible, mitigating this demands innovative analysis methods and collaboration between theoretical models like ours and data processing, with work on cataloging E-EMRIs/XMRIs being critical to informing these strategies, particularly in defining priors for the forest's properties. The challenge resembles debris removal in planetary radar or galactic foregrounds in CMB studies, where progress relies on iterative modeling and cross-validation.

Finally, we note that our analysis neglects the problem of the interplay between XMRIs and E-EMRIs and its influence on the forest’s spectral shape, neither LISA's modulation impact on the non-stationarity of the shape, and the role of extragalactic forests. Treating the forest’s morphology without taking into account these factors is an idealized approach, but as in the case of \cite{lynden1968gravo}, it is also our (other) justification that ``(...) only a fool tries the harder problem when he does not understand the simplest special case''.

\section*{Acknowldegments}

We thank Leor Barack, Neil Cornish and Scott Hughes for comments. This work has been partially supported by the National Natural Science Foundation of China (Grant No. 12247101), the Fundamental Research Funds for the Central Universities (Grant No. lzujbky-2024-jdzx06), the Natural Science Foundation of Gansu Province (No. 22JR5RA389) and the ``111 Center'' under Grant No. B20063. The support provided by China Scholarship Council (CSC) during a visit of Shao-Dong Zhao to the Universitat Politècnica de València is acknowledged.

\end{document}